\documentclass[conference]{IEEEtran}
\IEEEoverridecommandlockouts
\usepackage{cite}
\usepackage{amsmath,amssymb,amsfonts}
\usepackage{algorithmic}
\usepackage{graphicx}
\usepackage{textcomp}
\usepackage{xcolor}
\usepackage{color, colortbl}
\usepackage{framed}
\usepackage[skip=2pt]{caption}
\def\BibTeX{{\rm B\kern-.05em{\sc i\kern-.025em b}\kern-.08em
    T\kern-.1667em\lower.7ex\hbox{E}\kern-.125emX}}

\usepackage{draftwatermark}
\SetWatermarkText{ACCEPTED \\MANUSCRIPT}
\SetWatermarkScale{.6}
\SetWatermarkColor[gray]{0.9}
    
\begin{document}

\title{On the Impact of Entropy-based Features\thanks{This manuscript has been accepted for presentation at the IEEE International Symposium on Computers and Communications (ISCC 2026).}
}


\author{\IEEEauthorblockN{
Iuri Mundstock\IEEEauthorrefmark{1},
Abreu Quevedo\IEEEauthorrefmark{1},
Jéferson Campos Nobre\IEEEauthorrefmark{4},\\
Roben C. Lunardi\IEEEauthorrefmark{2},
Thiago L. T. da Silveira\IEEEauthorrefmark{3},
Bruno L. Dalmazo\IEEEauthorrefmark{1}
} 
\IEEEauthorblockA{\IEEEauthorrefmark{2}
IFRS and PUCRS, Porto Alegre, Brazil\\
E-mail: roben.lunardi@zonanorte.ifrs.edu.br}
\IEEEauthorblockA{\IEEEauthorrefmark{4}
Federal University of Rio Grande do Sul - UFRGS, Porto Alegre, Brazil\\
E-mail: jcnobre@inf.ufrgs.br}
\IEEEauthorblockA{\IEEEauthorrefmark{3}
Federal University of Santa Maria - UFSM, Santa Maria, Brazil\\
E-mail: thiagosilveira@inf.ufsm.br}
\IEEEauthorblockA{\IEEEauthorrefmark{1}Federal University of Rio Grande - FURG, Rio Grande, Brazil}%
E-mail: \{iurimundstock, abreu\_rg, dalmazo\}@furg.br
}

\maketitle

\begin{abstract}

Network anomaly detection is increasingly challenging due to the growing diversity and variability of traffic patterns, which are not always well captured by traditional statistical features. In this work, we explore the use of entropy as an additional feature to support supervised network traffic classification. The main idea is to use entropy to represent variability in selected traffic attributes, complementing conventional descriptors rather than replacing them. 
We integrate the entropy-based feature into a standard machine learning pipeline and evaluate its impact through a direct comparison between models trained with and without this feature. Experiments conducted on a public intrusion detection dataset show consistent improvements in classification performance, while the additional computational cost remains low. The analysis of confusion matrices indicates a reduction in misclassifications, especially in traffic scenarios with higher variability. Overall, the results suggest that entropy-based features offer a simple and practical way to enhance existing anomaly detection pipelines. This approach is particularly attractive in settings where lightweight feature engineering and interpretability are important, making entropy a useful complement to commonly used traffic features.
\end{abstract}


\section{Introduction}

The rapid evolution of distributed and cognitive cloud infrastructures, combined with the digitalization of urban environments, has led to the emergence of complex systems such as smart cities. These environments generate large volumes of high-dimensional and dynamic data, making efficient data representation and analysis a critical challenge. At the same time, the growing dependence on Internet-based services reinforces the importance of ensuring the security and reliability of network systems. According to the International Telecommunication Union, 5.5 billion people accessed the Internet in 2024, representing 68\% of the world’s population~\cite{quevedo2025choquet, quevedoRfXG}.

Among the main threats to these systems are denial-of-service (DoS) and distributed denial-of-service (DDoS) attacks, which can disrupt services and cause significant financial and operational damage. Therefore, the development of effective traffic analysis and anomaly detection mechanisms is essential to ensure system resilience~\cite{sbsegleite}. In addition to detection accuracy, a key challenge lies in efficiently representing network traffic data, as traditional approaches often rely on high-dimensional feature sets or handcrafted statistical descriptors that may fail to capture behavioral patterns.

Entropy-based representations have been widely used in network traffic analysis as an effective  way to capture variability and uncertainty~\cite{9689220, test-nemy}. However, most existing approaches are limited to static or coarse-grained formulations, typically applied to individual features or aggregated statistics. To overcome these limitations, this work proposes a temporal and multidimensional entropy modeling approach, capable of capturing both structural dependencies between traffic attributes and their evolution over time.

In the literature, several studies address DoS and DDoS detection using machine learning techniques~\cite{10195809,9350788}. However, many of these works rely on predefined feature sets and do not explore more expressive representations of traffic behavior. In this context, the proposed method advances the state of the art by introducing entropy-based features that improve anomaly detection performance while providing a more compact representation of high-dimensional data, particularly suitable for dynamic environments such as smart cities.


In this context, this work aims to develop a feature for network traffic classification based on entropy calculation. 
The main contributions of this work are:
(i) an empirical analysis of entropy-based features for network traffic anomaly detection;
(ii) an evaluation of their impact when combined with standard supervised classifiers;
(iii) a discussion of their applicability in resource-constrained detection pipelines.








The remainder of this paper is organized as follows: 
Section~\ref{literatura} presents a review of the current literature and discusses existing research gaps. Section~\ref{proposta} describes the step-by-step execution of the proposed approach, including the entropy calculation method and the process for generating a new feature. The analysis and discussion of the results are presented in Section~\ref{results}. Finally, Section~\ref{conclusao} concludes the paper and outlines directions for future research.

\section{Related work}
\label{literatura}

This section presents the systematic methodology used to retrieve existing studies in the context of distributed denial-of-service (DDoS) attack detection in computer networks using entropy and machine learning techniques. In this section, the research method adopted is described, as well as the articles identified through this methodology.

\subsection{Selection Method}

In the context of distributed denial-of-service (DDoS) attack detection, the keywords used for the literature search were “DDoS attack detection,” “Distributed Denial of Service attack detection,” “entropy,” and “machine learning.” Thus, the combination of keywords shown below was used to perform the query on the platform, and the criteria described in Table~\ref{tab:inclusion_exclusion_criteria} were applied, resulting in the set of studies presented in Table~\ref{tab:estudos}.


\begin{framed}
\centering
(``DDoS attack detection" OR ``Distributed Denial of Service attack detection") 
AND (``entropy" AND ``machine learning")
\end{framed}

\vspace{-0.2cm}

\begin{table}[htbp]
\centering
\caption{Inclusion (IC) and Exclusion (EC) Criteria}
\label{tab:inclusion_exclusion_criteria}
\begin{tabular}{l|l}

\textbf{Inclusion Criteria} &  \\
\hline
\rowcolor{lightgray} IC1 & Publication period between 2018 and 2025 \\

\hline
\textbf{Exclusion Criteria} &  \\

\rowcolor{lightgray} EC1 & No feature selection performed \\

EC2 & Does not use entropy \\

\end{tabular}
\end{table}

\vspace{-0.3cm}

\begin{table}[htbp]
\centering
\caption{Selected Studies}
\label{tab:estudos}
\begin{tabular}{l|c}

\textbf{Results} & \textbf{Total} \\
\hline
\rowcolor{lightgray} IEEE Xplore & 48 \\

Not included by IC1 & 02 \\

\rowcolor{lightgray} Excluded by EC1 & 21 \\

Excluded by EC2 & 05 \\

\rowcolor{lightgray} \textbf{Selected Articles} & \textbf{20} \\

\end{tabular}
\end{table}

\vspace{-0.25cm}

\begin{table}[htbp]
\vspace{0.3cm}
\caption{Related Works Using Entropy}
\label{tab:related_works_ieee}
\centering
\footnotesize
\begin{tabular}{p{0.6cm}|p{2.2cm}|p{3.6cm}}

\textbf{Ref.} & \textbf{Dataset} & \textbf{Main Contribution} \\

\hline
\rowcolor{lightgray}
\cite{9786423} & CIC-IDS17 
& RF-based DDoS detection using entropy features. \\

\hline
\cite{10284799} & CIC-DDoS19 
& Entropy-based information gain for feature ranking. \\

\hline
\rowcolor{lightgray}
\cite{9800775} & MTS-IoT 
& Supervised entropy-driven DDoS detection framework. \\

\hline
\cite{8796637} & CIC-IDS17 
& Neural network enhanced with entropy features. \\

\hline
\rowcolor{lightgray}
\cite{10026905} & CAIDA 
& Packet size entropy for binary traffic classification. \\

\hline
\cite{10741358} & CIC-IDS17, CIC-DDoS19 
& Deep learning with entropy features and noise regularization. \\

\hline
\rowcolor{lightgray}
\cite{8765599} & Own 
& Distributed detection using entropy, ML, and fuzzy logic. \\

\hline
\cite{9186014} & NSL-KDD 
& K-means and KNN with entropy-based features. \\

\hline
\rowcolor{lightgray}
\cite{10464280} & -- 
& Survey of entropy-based DDoS detection methods. \\

\hline
\cite{9350788} & LLDoS1.0, 2.0.1 
& Entropy-based feature selection with RF classifier. \\

\hline
\rowcolor{lightgray}
\cite{10227717} & NSL-KDD 
& SVM combined with entropy features. \\

\hline
\cite{9817297} & BoT-IoT 
& Feature selection for entropy-based detection. \\

\hline
\rowcolor{lightgray}
\cite{10195809} & Own 
& RF using entropy measures and Gini coefficient. \\

\hline
\cite{9738866} & Own 
& CNN-based detection using relative entropy. \\

\hline
\rowcolor{lightgray}
\cite{9539920} & Own 
& RF-based detection using Gini impurity. \\

\hline
\cite{9090824} & CIC-DDoS19 
& FlowGuard: entropy-based IoT DDoS filtering. \\

\hline
\rowcolor{lightgray}
\cite{9689220} & CIC-DDoS19, MAWI 
& Entropy-driven traffic patterns via HTM. \\

\hline
\cite{9683214} & CSE-CIC18, CIC-DDoS19 
& Anomaly detection combining entropy and ML. \\


\end{tabular}
\end{table}

\subsection{Discussion}

As shown in Table~\ref{tab:related_works_ieee}, many studies make use of datasets provided by the Canadian Institute for Cybersecurity, such as CIC-IDS2017 and CIC-DDoS2019, available in~\cite{datasets}. However, there are differences between these datasets, as each one contains distinct distributions of attack types. The dataset selected for this research is CIC-IDS2017, which has been used in studies such as~\cite{10741358}, \cite{8796637}, and \cite{9786423} for the evaluation of the proposed methods.

Several studies focus on selecting the most relevant features for detecting denial-of-service attacks using different selection methods. Among them, entropy stands out because it has shown promising results, which is why it is also adopted in this work. In addition, many machine learning techniques have been applied to this problem, with Random Forest being one of the most frequently cited models. As an ensemble of decision trees, it has consistently demonstrated strong performance in attack detection.

Finally, a review of the literature reveals that although many studies focus on detecting and mitigating distributed denial-of-service attacks using entropy, none of them primarily emphasize the generation of rules for blocking malicious traffic. Therefore, this work employs a Random Forest model (widely recognized in the literature as an effective method for real-time traffic classification) with the objective of generating rules that enable the blocking of malicious traffic.
\section{Proposal}
\label{proposta}

This section presents the proposal of the work and describes, step by step, how its implementation is carried out. The objective of this work is to develop a feature based on entropy for anomaly detection and to validate it using a Random Forest model, classifying network traffic as legitimate or malicious.

The underlying hypothesis of the proposed feature is that anomalous traffic exhibits higher structural unpredictability in selected packet attributes when compared to benign traffic. Such unpredictability is not necessarily reflected in first-order statistics, motivating the use of entropy as a complementary descriptor.

\subsection{Conceptual model}

\begin{figure}[htbp]
\centering
\includegraphics[width=0.45\textwidth]{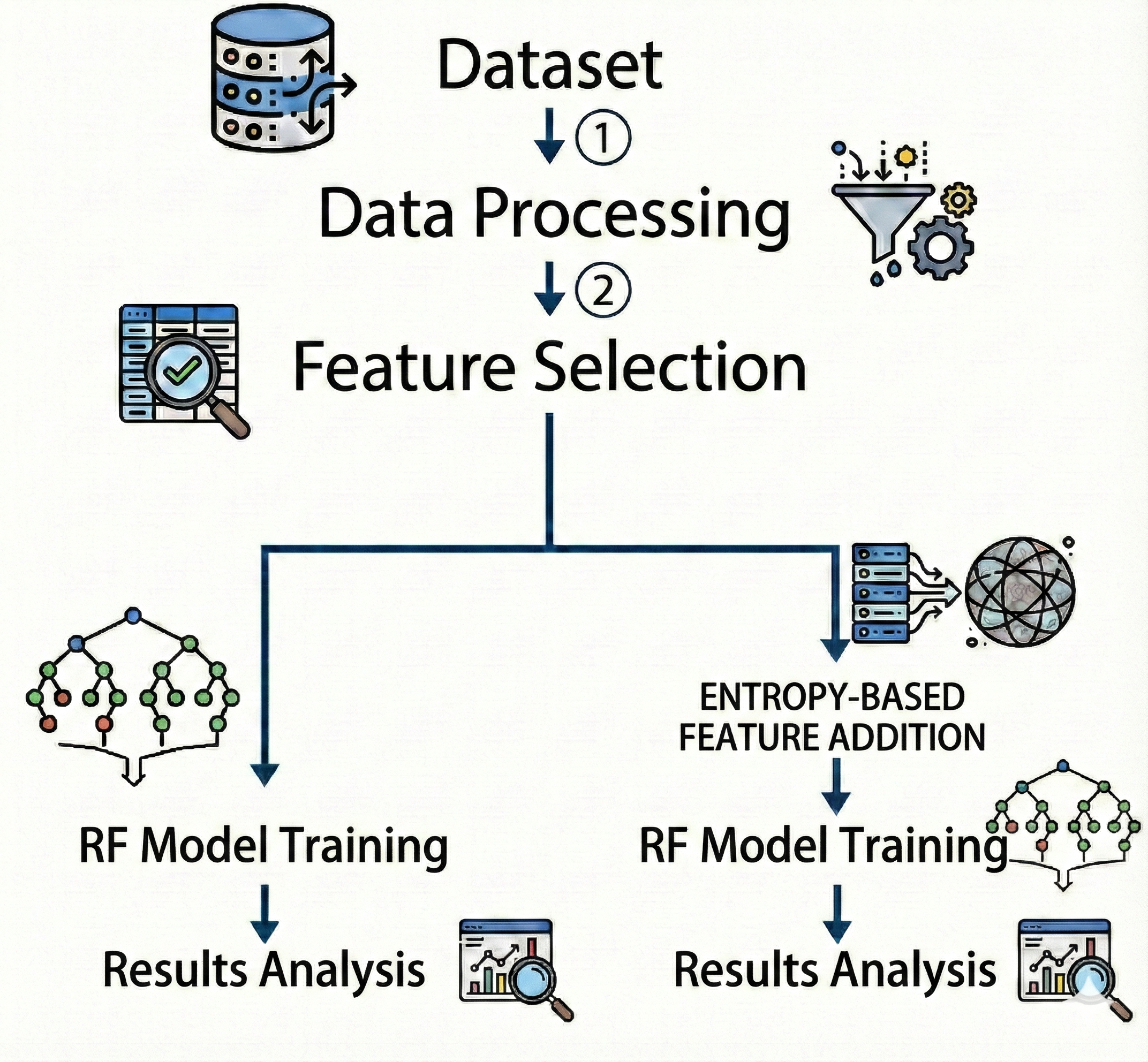}
\caption{Conceptual model}
\label{fig:Organograma}
\end{figure}

In the first step, as shown in Fig.~\ref{fig:Organograma}, a preprocessing procedure and a time series analysis were applied to the dataset with the objective of understanding and transforming the data to facilitate its use in the subsequent steps. This stage is essential to ensure that the data are properly treated and ready to be efficiently used in later analyses.

The dataset is divided into training (70\%) and testing (30\%) sets. Entropy is computed independently for each subset to avoid data leakage. A Random Forest model is then trained using the enriched feature set and evaluated on the test data.

\vspace{-0.2cm}

\vspace{-0.2cm}

Unlike approaches that rely on complex representations or deep feature learning, this work explores entropy as a lightweight statistical proxy for capturing dispersion patterns in network traffic attributes. The underlying hypothesis is that anomalous traffic often exhibits irregular or highly concentrated distributions over selected feature dimensions, which can be effectively summarized through entropy-based measures.

Then, with the dataset properly divided, the entropy is calculated individually for each part. This calculation is used to enable the Random Forest model to better understand network behavior patterns, aiming to improve its performance.

After that, the Random Forest model is trained for traffic classification. Random Forest was selected due to its robustness to noise, ability to handle heterogeneous feature scales, and effectiveness in modeling non-linear decision boundaries without extensive hyperparameter tuning. This makes it a suitable baseline for evaluating the impact of feature engineering strategies, particularly the inclusion of entropy-based features. 

However, it is important to note that Random Forest presents limitations in terms of scalability and sensitivity to class imbalance in real-time network environments. Despite these constraints, it provides a reliable and widely adopted reference model for assessing whether entropy-derived features contribute complementary discriminative information.


Subsequently, a traffic simulation is carried out using the second portion of the dataset, representing thirty percent of the original data, to perform classification using the Random Forest model trained in the previous step. Based on the results of this classification, rules are generated to block malicious traffic.

Finally, the results are presented in detail, including analyses, interpretations, and graphical representations of the outcomes, as well as the application of the Random Forest classifier and the evaluation of anomaly detection. The strengths and limitations of the proposed method are discussed, along with insights and recommendations for future research in this area. This stage aims to provide a comprehensive and well-founded overview of the study’s results, contributing to the advancement of knowledge in anomaly detection in computer networks.

\vspace{-0.1cm}

\subsection{Dataset}

This work uses the CIC-IDS2017 dataset, which contains network traffic records collected over five consecutive days, including both benign behavior and a diverse set of attack scenarios. The complete dataset is considered in order to provide a comprehensive evaluation and improve the generalizability of the results. It is available in CSV format, with a total size of 1725.05 MB and 2,830,743 input instances.

The dataset includes multiple types of attacks distributed across different days, such as brute force (FTP and SSH), several DoS variants (e.g., Hulk, Slowloris, GoldenEye), web-based attacks (e.g., XSS and SQL injection), infiltration, and botnet activity. These scenarios occur at different time intervals and under varying traffic conditions, enabling the evaluation of the proposed method in heterogeneous and dynamic environments. It is also worth noting that the first day (Monday, 07/03/2017) contains only benign traffic and is included to represent normal network behavior.

\vspace{-0.1cm}
\section{Evaluation and Results}
\label{results}

This section presents the results obtained in this research, including the graphical representation of the experiments and a detailed discussion of the findings. The performance of each model is analyzed based on the adopted evaluation metrics, allowing a fair comparison between the proposed approaches. To ensure statistical robustness and reduce variability due to data partitioning, all experiments were conducted using 15-fold stratified cross-validation, and the reported results correspond to the mean performance across the folds. All classifiers were evaluated using the default decision threshold provided by the learning algorithm, and no threshold tuning or class-dependent calibration was performed in any approach.

\subsection{Entropy calculation}

In this work, entropy is not computed as a static descriptor over individual features, but rather as a dynamic and multidimensional representation of network traffic behavior.

Given a discrete random variable \( X \), the entropy is defined:

\[
H(X) = - \sum_{i=1}^{n} p(x_i) \log_b p(x_i)
\]

where:

\begin{itemize}

  \item \( H(X) \): represents the \textbf{entropy} of the variable \( X \), measuring the uncertainty or unpredictability associated with its outcomes;
  
  \item \( p(x_i) \): denotes the \textbf{probability} of occurrence of the event \( x_i \);
  
  \item \( \log_b \): is the logarithm with base \( b \), where \( b = 2 \) is used to express entropy in \textbf{bits}.
  
\end{itemize}

Unlike conventional approaches, which compute entropy over single attributes, this work employs a \textbf{multidimensional formulation} based on the joint distribution of multiple traffic features. In this case, the entropy is computed as:

\[
H(X, Y) = - \sum_{x \in X} \sum_{y \in Y} p(x, y) \log p(x, y)
\]

where \( p(x, y) \) represents the joint probability of observing the pair \( (x, y) \), enabling the capture of structural relationships between traffic attributes.

Furthermore, to model the temporal dynamics of network traffic, entropy is computed over \textbf{sliding windows} of size \( w \). For each time step \( t \), the entropy \( H_t \) is calculated based on the subset of observations within the interval \( [t-w, t] \). From this temporal sequence, additional features are derived, including:

\begin{itemize}

  \item \textbf{Entropy variation}:
  \[
  \Delta H_t = H_t - H_{t-1}
  \]

  \item \textbf{Local mean entropy}:
  \[
  \mu_t = \frac{1}{w} \sum_{i=t-w}^{t} H_i
  \]

  \item \textbf{Entropy volatility}:
  \[
  \sigma_t = \sqrt{\frac{1}{w} \sum_{i=t-w}^{t} (H_i - \mu_t)^2}
  \]

\end{itemize}

These extensions allow the proposed method to capture both the \textbf{structural dependencies} between traffic attributes and their \textbf{temporal evolution}, providing a more expressive representation of anomalous behavior when compared to traditional entropy-based features.

\vspace{-0.1cm}

\subsection{Random Forest and classification}

Two versions of the dataset were considered: one composed of the original features after the application of the preprocessing stage, and another enriched with the entropy values calculated based on the formulation presented earlier. From these two configurations, the classification stage was carried out, enabling a comparative analysis of the impact of the entropy-based feature on the model’s performance.

Thus, for the implementation, the Python programming language was adopted due to its wide use in machine learning applications. The Pandas library was employed for dataset loading and manipulation, Matplotlib for result visualization, NumPy for numerical operations, and Scikit-learn (\texttt{sklearn}) for data preprocessing and the application of machine learning algorithms, including the Random Forest model used in this work, which facilitated the implementation and the presentation of the results.




    

\vspace{-0.1cm}

\subsection{Analysis of Results}

In this section, the results obtained from the conducted experiments are presented, with the objective of evaluating the impact of the entropy-based feature on the performance of the network traffic classification model. We acknowledge that the use of a single dataset may limit the generalization of the reported results. The goal of this evaluation is not to claim universal applicability, but to assess whether entropy-based features provide consistent gains within a controlled and widely used benchmark.

\begin{figure}[htb]
\centering
\includegraphics[width=0.25\textwidth]{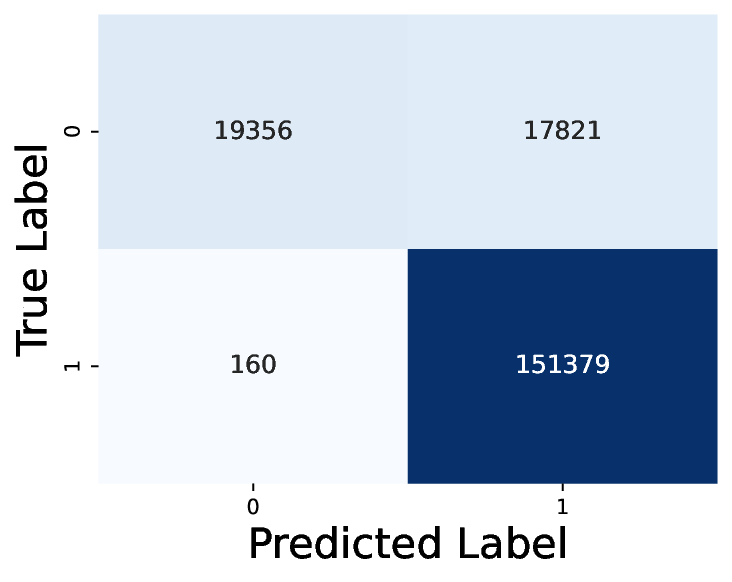}
\caption{Baseline confusion matrix}
\label{fig:matrizconfusao1}
\end{figure}

\begin{figure}[htb]
\centering
\includegraphics[width=0.25\textwidth]{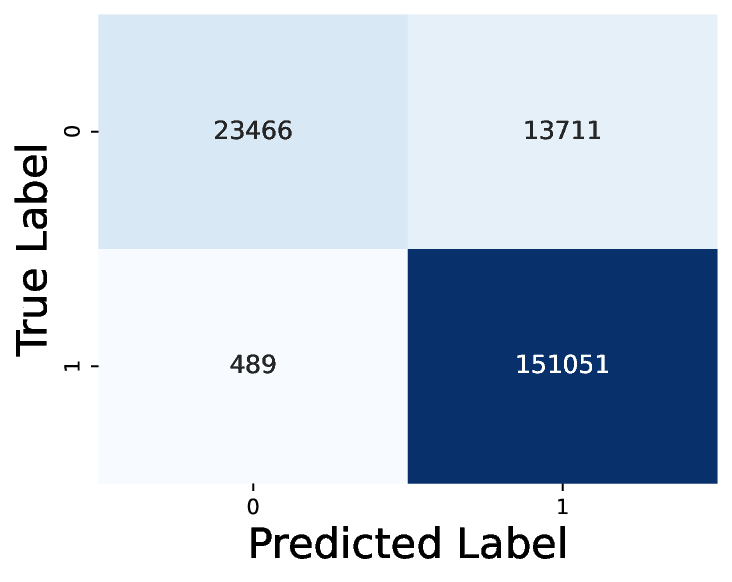}
\caption{Confusion matrix with entropy-based feature}
\label{fig:matrizconfusao2}
\end{figure}

Initially, two confusion matrices are presented for comparison of the classification performance: the first, illustrated in Figure \ref{fig:matrizconfusao1}, corresponds to the baseline scenario without the introduction of the entropy-based feature, while the second, shown in Figure~\ref{fig:matrizconfusao2}, refers to the scenario with the inclusion of the proposed feature.

\vspace{-0.3cm}

\vspace{-0.5cm}

\vspace{-0.2cm}

The confusion matrix corresponding to the baseline scenario (without the inclusion of entropy-based features) indicates that the model exhibits a high capability for detecting malicious traffic, as evidenced by the high number of true positives (151,379) and the low number of false negatives (160). This behavior is desirable in intrusion detection systems, as it minimizes the occurrence of undetected attacks. However, a considerable number of false positives (17,821) are observed, which implies the incorrect classification of legitimate traffic as malicious. In real-world environments, this may lead to unnecessary alerts and potential degradation of network service availability and quality.

With the introduction of the entropy-based features, the confusion matrix reveals a reduction in the number of false positives, decreasing to 13,711, indicating an improvement in the model’s ability to distinguish legitimate traffic patterns from anomalous ones. Additionally, the number of true negatives increases to 23,466, reinforcing this improvement. On the other hand, an increase in the number of false negatives (489) is observed, indicating that a portion of malicious traffic is no longer detected. Despite this, the model still maintains a high number of true positives (151,051), preserving strong detection capability. These results highlight a trade-off between reducing false positives and increasing false negatives, demonstrating that the proposed approach improves classification reliability while slightly impacting detection sensitivity.

\begin{figure}[htb]
\centering
\includegraphics[width=0.48\textwidth]{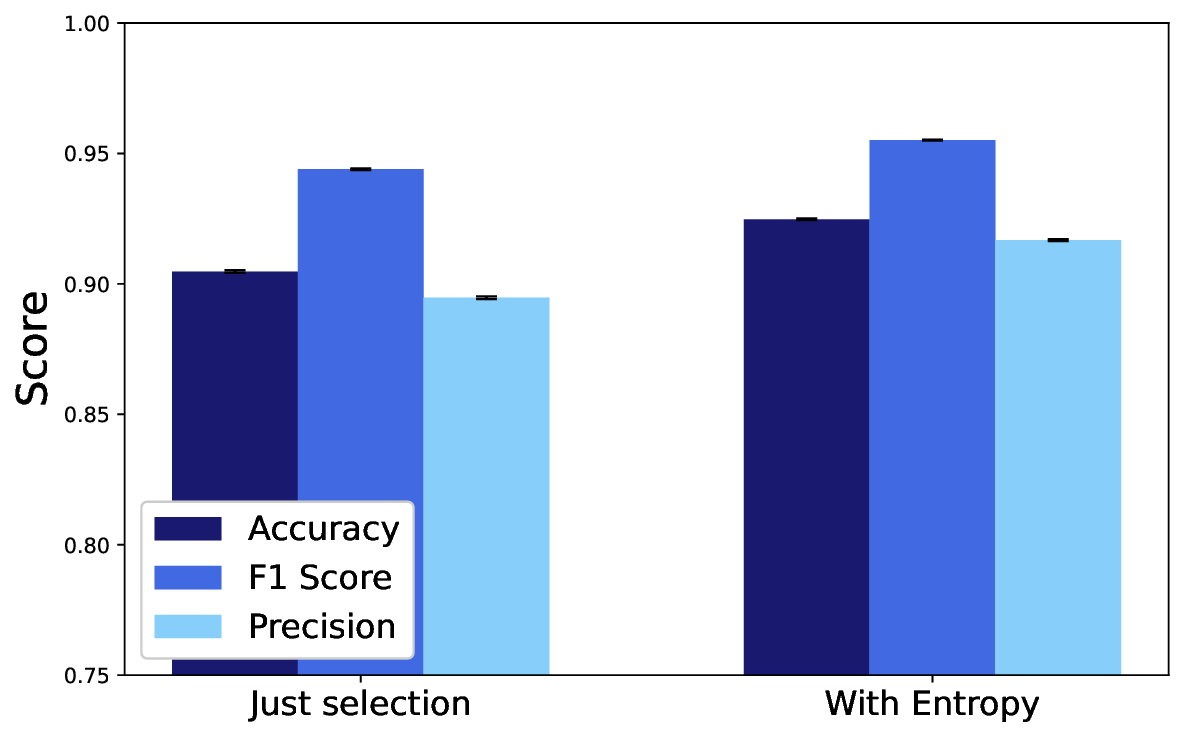}
\caption{Comparative performance with and without entropy-based feature}
\label{fig:comprativo}
\end{figure}

Next, Figure \ref{fig:comprativo} presents a comparison of the performance of the network traffic classification model in scenarios with and without the inclusion of entropy-based features, considering the metrics of accuracy, F1-score, and precision. In the baseline scenario, where only feature selection is applied, the model achieves an accuracy of 0.9047±0.0005, an F1-score of 0.9439±0.0003, and a precision of 0.8947±0.0005, indicating strong performance in detecting malicious traffic. However, the relatively lower precision, combined with a higher false positive rate (0.4794±0.0026), highlights the presence of a considerable number of incorrect classifications of legitimate traffic as malicious, which may compromise system reliability in real-world environments.

\vspace{-0.3cm}

\vspace{-0.2cm}

With the inclusion of the entropy-based features, consistent improvements are observed across the evaluated metrics, with accuracy reaching 0.9248±0.0003, the F1-score 0.9551±0.0002, and precision 0.9168±0.0004. The increase in precision, along with the reduction in the false positive rate to 0.3688±0.0018, indicates a substantial improvement in the model’s ability to distinguish legitimate traffic from anomalous behavior, corroborating the observations from the confusion matrix analysis. At the same time, a slight reduction in recall (0.9968±0.0001 compared to 0.9989±0.0001) and an increase in the false negative rate are observed, reflecting a trade-off between reducing false alarms and maintaining maximum detection sensitivity.

Overall, although the improvements are not uniform across all performance aspects, the results demonstrate that the proposed approach enhances the model’s discriminative capability. In particular, the reduction in false positives, combined with the maintenance of a high detection rate, highlights the effectiveness of entropy-based features in improving the balance between reliability and detection performance, which is critical for practical anomaly detection systems in network environments.

\begin{figure}[htb]
\centering
\includegraphics[width=0.40\textwidth]{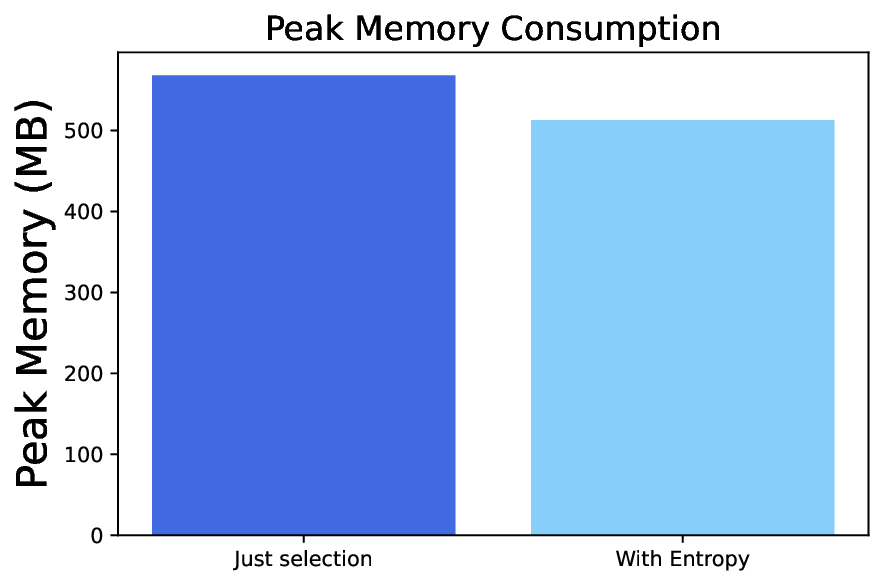}
\caption{Peak memory consumption observed during execution}
\label{fig:memorypeak}
\end{figure}

\begin{figure}[htb]
\centering
\includegraphics[width=0.40\textwidth]{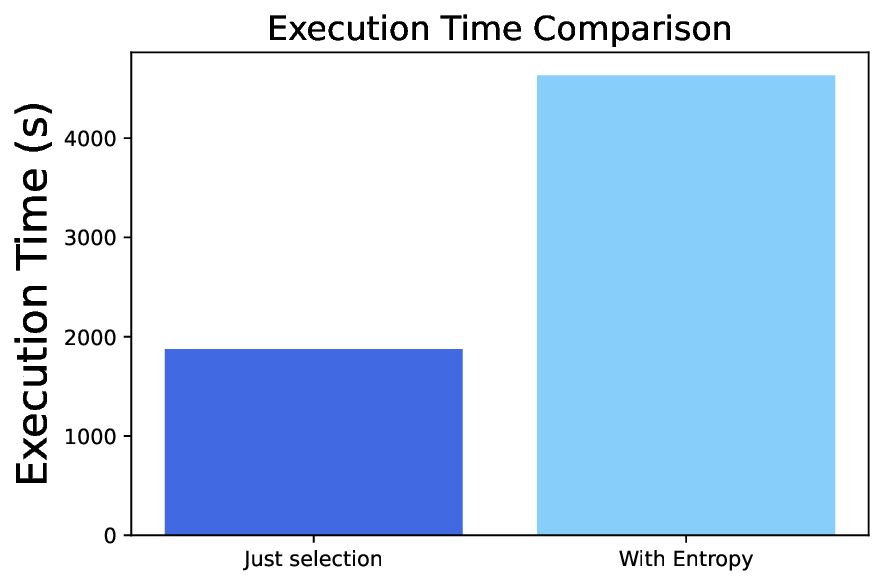}
\caption{Training and testing time}
\label{fig:trainningtime}
\end{figure}

Finally, Figures \ref{fig:memorypeak} and \ref{fig:trainningtime} compare memory usage and execution time between the evaluated scenarios. The inclusion of entropy-based features leads to an increase in execution time, rising from 1873.86 seconds in the baseline to 4632.05 seconds in the proposed approach. In terms of memory, the results show a different behavior: while the baseline presents a peak memory usage of 568.21 MB, the entropy-based approach reaches a slightly lower peak of 512.87 MB.

\vspace{-0.3cm}

\vspace{-0.5cm}

\vspace{-0.2cm}

The increase in execution time reflects the additional computational cost associated with the calculation of temporal and multidimensional entropy features, particularly due to the use of sliding windows and joint probability distributions. As a result, the proposed method is more computationally demanding in terms of processing time compared to the baseline approach.

On the other hand, this additional cost is accompanied by improved classification performance, particularly in terms of precision and overall balance between false positives and false negatives. Therefore, the results suggest that the proposed method represents a trade-off between computational efficiency and detection effectiveness.

From a practical perspective, this trade-off may be acceptable in scenarios where detection reliability is prioritized over execution time, such as offline analysis or high-capacity environments. However, in real-time or latency-sensitive systems, further optimizations may be required to reduce the computational overhead associated with entropy calculations.

Overall, the observed reduction in false positives, combined with the preservation of high detection capability, highlights the practical value of the proposed approach. These findings reinforce that temporal and multidimensional entropy features provide a more expressive representation of network traffic, enabling improved anomaly detection at the cost of increased processing time.

\vspace{-0.1cm}

\section{Final Considerations}
\label{conclusao}

This work addresses the challenge of representing high-dimensional and dynamic network data in complex environments such as smart cities, investigating the use of temporal and multidimensional entropy-based feature engineering to enhance network traffic anomaly detection. By modeling both structural relationships and temporal dynamics of traffic attributes, the proposed approach extends traditional entropy formulations and provides a more expressive representation of network behavior, albeit at a higher computational cost.

Experimental results demonstrate that the proposed method improves the balance between false positives and false negatives, reducing false alarms while maintaining a high detection capability. This behavior is particularly relevant in practical intrusion detection systems, where excessive false positives may compromise usability and operational efficiency. These results highlight that the proposed approach should be interpreted as a performance-oriented enhancement rather than a lightweight solution.

However, the results also show that these improvements come at the cost of increased computational cost, particularly in terms of execution time, due to the additional processing required for entropy computation over sliding windows and joint distributions. This highlights an inherent trade-off between detection performance and computational efficiency, which must be considered depending on the target application scenario, especially in large-scale or time-sensitive environments.

Overall, the findings indicate that entropy-based representations, when extended to capture temporal and multidimensional characteristics, can significantly enhance anomaly detection performance while providing a richer characterization of complex traffic patterns. Future work will focus on optimizing the computational efficiency of the proposed approach, as well as evaluating its applicability in real-time, distributed, and large-scale environments. Additionally, further investigation into adaptive mechanisms and integration with explainability techniques may contribute to more robust and interpretable intrusion detection systems.

\bibliography{bibliografia}
\bibliographystyle{ieeetr}

\end{document}